\documentclass[lettersize,journal]{IEEEtran}
\usepackage{amsmath,amsfonts}
\usepackage{algorithmic}
\usepackage{array}
\usepackage[caption=false,font=normalsize,labelfont=sf,textfont=sf]{subfig}
\usepackage{textcomp}
\usepackage{stfloats}
\usepackage{url}
\usepackage{verbatim}
\usepackage{graphicx}
\usepackage{amssymb}
\usepackage{bbm}
\usepackage{boldline} 
\def\BibTeX{{\rm B\kern-.05em{\sc i\kern-.025em b}\kern-.08em
    T\kern-.1667em\lower.7ex\hbox{E}\kern-.125emX}}
\usepackage{balance}
\usepackage{cite}
\begin{document}
\title{Targeted Analysis of High-risk States Using an Oriented Variational Autoencoder}
\author{Chenguang Wang,~\IEEEmembership{Graduate Student Member,~IEEE}, Ensieh Sharifnia, \\Simon H. Tindemans,~\IEEEmembership{Member,~IEEE}, and Peter Palensky,~\IEEEmembership{Senior Member,~IEEE}

\thanks{The authors are with the Department of Electrical Sustainable Engineering, Delft University of Technology, 2628 CD, Delft, The Netherlands (emails: \{c.wang-8, e.sharifnia, s.h.tindemans, p.palensky\}@tudelft.nl).}}

\markboth{}%
{Targeted Analysis of High-Risk States Using an Oriented Variational Autoencoder}

\maketitle

\begin{abstract}

Variational autoencoder (VAE) neural networks can be trained to generate power system states that capture both marginal distribution and multivariate dependencies of historical data. The coordinates of the latent space codes of VAEs have been shown to correlate with conceptual features of the data, which can be leveraged to synthesize targeted data with desired features. However, the locations of the VAEs' latent space codes that correspond to specific properties are not constrained. Additionally, the generation of data with specific characteristics may require data with corresponding hard-to-get labels fed into the generative model for training. In this paper, to make data generation more controllable and efficient, an oriented variation autoencoder (OVAE) is proposed to constrain the link between latent space code and generated data in the form of a Spearman correlation, which provides increased control over the data synthesis process. On this basis, an importance sampling process is used to sample data in the latent space. Two cases are considered for testing the performance of the OVAE model: the data set is fully labeled with approximate information and the data set is incompletely labeled but with more accurate information. The experimental results show that, in both cases, the OVAE model correlates latent space codes with the generated data, and the efficiency of generating targeted samples is significantly improved.
\end{abstract}


\section{Introduction}
\IEEEPARstart{T}{he} analysis of power system performance across a large range of representative scenarios is of great significance for power system planning and risk assessments \cite{bloomfield2021quantifying,panciatici2012operating,wang2019risk}. However, data scarcity, reluctance to share, and confidentiality concerns may limit the amount of available historical data, which is a key source of representative scenarios. To this end, it is highly desirable to have access to generative models that produce limitless non-repeating data, reproducing both univariate distributions and inter-dependencies observed in historical data \cite{Konstantelos2019}.

In recent years, with the development of deep neural network-related technology, a promising data-driven-based approach has been proposed in the form of \emph{variational autoencoders} (VAEs)\cite{kingma2013auto,doersch2016tutorial, Kingma2019}. These networks encode high-dimensional historical data (observations) to a latent code in a lower-dimensional \emph{latent space} - and reconstruct similar observations from this code. Features of historical data are learned, so that novel but realistic data points can be created from random codes. In recent research, the VAE model has been successfully used in generating electricity load profiles \cite{pan2019data}. Similarly, Generative Adversarial Networks (GANs) have been used to generate realistic power system states \cite{baasch2021conditional,wang2020generating}, but such models do not provide straightforward access to latent representations \cite{creswell2018generative}.

Coordinates of latent space codes of VAEs have been shown to correlate with conceptual features of the data \cite{spinner2018towards,way2018extracting}. These coordinates can then be used to synthesize targeted data with desired features \cite{sarkar2021generating} (e.g., game scenarios). Although the degree of informativeness and orthogonality of latent space variables can  be influenced by the training process \cite{burgess2018understanding}, the interpretation of individual latent variables and the existence of particular concepts are not determined \emph{a priori}. 

For power system applications, it is often valuable to generate samples that pertain to certain operating conditions. One use case is when performing studies for a particular time window, geographical area or otherwise clearly delineated set of conditions. In this case, a  generative model can be conditioned on the selection criterion of interest, using e.g. the Conditional VAE (CVAE) instead of the regular VAE. This was done, for example, in \cite{wang2022generating}, for country-level load snapshots conditioned on the hour-of-day and in \cite{wang2022ISGTgenerating} for 24-hour load profiles of industrial users conditioned on the month-of-year. 

A particularly important use case for targeted sampling occurs in power system risk assessment. As power systems are usually highly reliable, unbiased (Monte Carlo) sampling of states results in excessively high sample count requirements. The accurate estimation of risks can be sped up using importance sampling (IS) \cite{799896, 7581049,ALMEIDA2021107001} 
, which samples high-impact states more often and compensates for the resulting bias by adjusting sample weights. However, this combines the challenge of targeted sampling with (1) knowing \emph{which} states to target and (2) calculating the correct sample weights. 

One approach is to use a bottom-up model to generate states and change its parameters for optimal importance sampling. 
This approach is used, for example, in \cite{da2010generating}, where generator forced outage rates are modified to speed up generation adequacy assessment, and the cross-entropy method is used to optimize model parameters in a number of stages. This approach can lead to very high speedups 
, but it requires the availability of a bottom-up generative model and some degree of expert knowledge about which parameters to modify.

Subset simulation represents an alternative approach, where regions of interest in state space are identified and refined in iterations \cite{6845377}. 
This requires learning the region of interest and generating states according to this region of interest, ideally without resorting to a sample filtering (accept-reject) approach. A particular challenge is that, to avoid biasing the risk estimate, it is essential that no states with a non-zero impact are excluded from sampling. Moreover, typically no distinction is made between the likelihood of sampling low-impact and high-impact states. 

In this paper, we address the challenges identified above by proposing the Oriented Variational Autoencoder (OVAE), a data-driven generative model. Compared to the regular (C)VAE model, it provides increased control over the data synthesis process by correlating the features of interest derived from inputs and the data encoded in the latent space. The model can naturally be used for importance sampling, and can also be trained on incomplete labels. The main contributions of this paper are as follows:

\begin{itemize}

	\item We propose the oriented variational autoencoder (OVAE) that maximises the Spearman correlation of one latent dimension with a feature of interest.

    \item We demonstrate that an OVAE model can be trained in a semi-supervised manner using partially labeled data, which is essential when the process to labels is computationally expensive.
	
	\item Through comprehensive experiments, we test the performance of the OVAE-based generator and its ability to generate calibrated biased samples.
 
    \item The effectiveness of the OVAE model in importance sampling applications is investigated in a case study of multi-area adequacy assessment.

\end{itemize}

\section{Data Generation Mechanism}{\label{sec:Mechanism}}

In this section, a novel multivariate data generation mechanism is proposed, based on the \emph{oriented variational autoencoder} (OVAE). Importance sampling for system adequacy assessment is briefly reviewed in Section \ref{IS_for_Risk}, and the standard variational autoencoder formulation is summarized in Sections \ref{sec:naive_VAE} and \ref{sec:naive_VAE_function}. The Oriented VAE and its use in importance sampling are explained in Sections \ref{OVAE_model} and \ref{sec:IS-OVAE}, respectively. 

\subsection{Importance Sampling for Risk Assessment} \label{IS_for_Risk}

Quantitative risk assessment for power systems aims to compute one or more numerical risk indices. Often, these are long-run expectations of an operational performance measure, i.e., $r = \mathbb{E}_X[h(X)]$. For example, popular metrics for system adequacy assessment are \emph{Loss Of Load Expectation} (LOLE [h/year]) and \emph{Expected Energy Not Served} (EENS [MWh/year]): LOLE (as measured per hour, also known as LOLH) is the expected number of hours per year during which the supply does not meet demand; EENS is the expected amount of energy demand per year that cannot be supplied. Monte Carlo simulations can be used to estimate such risks by randomly selecting power system states $x$ (indexed by $i$) according to their probability density $p(x)$ and calculating the average of the impact $h(x)$ over all sampled states as
\begin{equation}
    \label{eq:mc_risk}
    \hat{r}_{MC} = \frac{1}{m}\sum_{i=1}^{m} h(x_i).
\end{equation}
However, random sampling is computationally inefficient for highly reliable systems when only a small fraction of states contribute, i.e. when $h(x)=0$ for most states $x$.

Importance sampling \cite{roy2013reliability} 
changes the sampling probability distribution to preferentially select samples with higher impact. This reduces the variance of the estimator, and therefore estimates risk values more accurately than the regular Monte Carlo method, for the same number of samples. When states $x'_i$ are sampled according to the modified distribution $q(x)$, the risk is estimated as 
\begin{equation}
    \label{eq:is_risk}
    \hat{r}_{IS} = \frac{1}{m}\sum_{i=1}^{m} h(x'_i)w(x'_i), 
\end{equation}
with sampling weights $w(x)=\frac{p(x)}{q(x)}$ that ensure an unbiased estimate despite the biased sampling procedure. It is easy to verify that optimal performance is attained when the sampling distribution $q(x)$ is chosen as 
\begin{equation} \label{eq:is-optimal}
    q^*(x)=\frac{h(x)p(x)}{E_{X\sim p(x)}[h(X)]}=\frac{h(x)p(x)}{r}.
\end{equation}
In fact, this choice reduces the variance to zero, and only a single sample is required. Of course, this distribution is unattainable in practice, as it depends on the quantity-to-be-estimated $r$ in the denominator.

Implementing an effective importance sampling procedure requires the following elements:
\begin{enumerate}
    \item A sufficiently flexible model that generates samples $X \sim q(x; \theta)$, where $\theta$ represents parameters that control the sampling distribution. 
    \item Knowledge of the sample impact distribution $h(x)$, so that the generative model can be tuned accordingly.
    \item An expression for the likelihood ratio (sample weight) $p(x)/q(x)$.
\end{enumerate}
However, in the real world, these requirements are often not met: (1) Common parametric distributions may not be sufficiently flexible to capture complex data distributions. (2) The evaluation of impacts may be computationally expensive. (3) Many generative models do not have an expression for the likelihood ratio, compared to the baseline model. In section \ref{sec:IS-OVAE}, we will demonstrate that the OVAE model, introduced below, provides a natural framework to address all these challenges.

\subsection{VAE-Based Generative Model} \label{sec:naive_VAE}

The \emph{variational autoencoder} (VAE) is a neural network architecture that is trained to learn the salient features of historical data by mapping (\emph{encoding}) historical system states onto a lower-dimensional latent space where the latent distribution is approximately multi-dimensional independent normal - and transforming latent vectors back (\emph{decoding}) into a high-dimensional state space \cite{Kingma2019}. The decoder is then used to generate representative states by decoding the latent vector. Consequently, the model is able to generate samples with a similar distribution to the historical data. We note that the latent (i.e., hidden) representation of a data point is used solely to facilitate reconstruction and synthesis. It does not need to be imbued with a particular meaning.
\begin{figure*} 
  \centering 
    \includegraphics[scale=0.56]{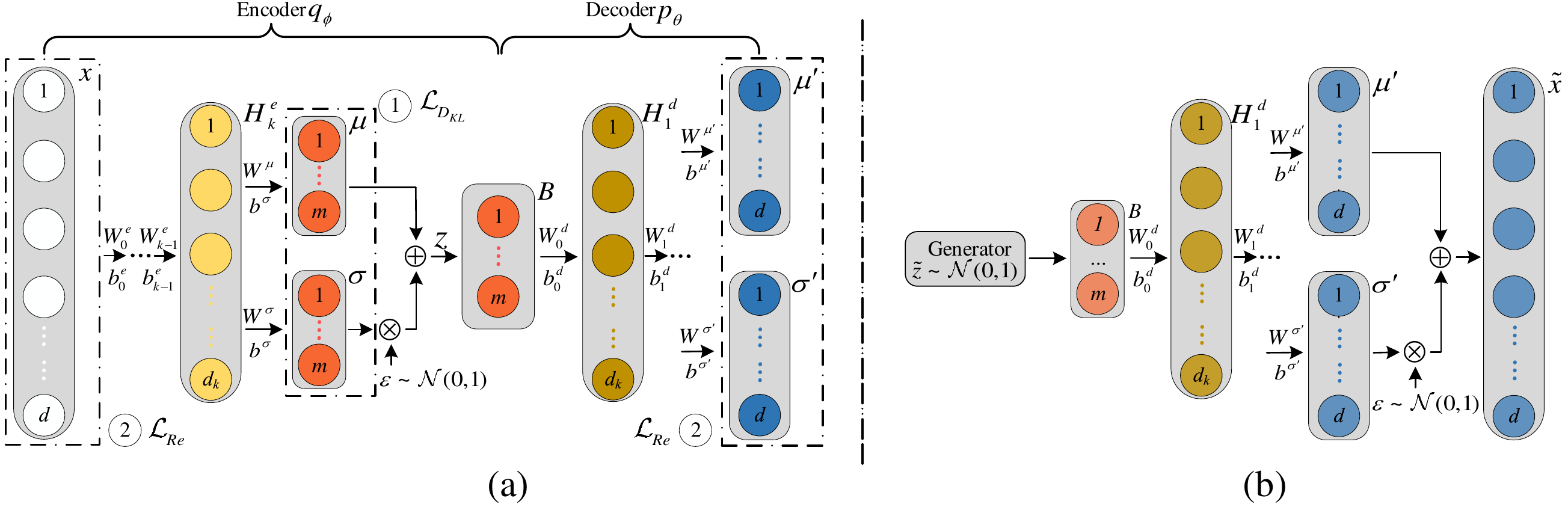}
  \caption{Schematic of a VAE model. (a) The structure of a VAE model for training. (b) The structure of a VAE model when it's utilized as a generator.}
  \label{fig:VAE}  
\end{figure*}

In the training process, the specific structure of the VAE algorithm is depicted in Fig.~\ref{fig:VAE} (a). The \emph{encoder} maps the $d$-dimensional input data $x$ to the code $z$ in the lower-dimensional latent space through $k$ hidden layers $H_l^e$, $l=1,\ldots,k$. Weight matrices $W_{l}^{e}$ and bias vectors $b_{l}^{e}$ are utilized in the encoding process as
\begin{subequations}
		\begin{align}\label{eq:y}
		\left(\begin{matrix} \mu \\ \sigma \end{matrix}\right)  = & \left(\begin{matrix} 	
		W^{\mu} \\ W^{\sigma} \end{matrix}\right)(a(W_{k}^e(\ldots 	
		a(W_{1}^{e}x + b_{1}^{e})\ldots)+b_{k}^e)) \nonumber \\ &
		+ \left(\begin{matrix} b^{\mu} 	\\ b^{\sigma} \end{matrix}\right)\,,
		\\
	\label{eq:hatz}
	z =& \mu(x)+ \epsilon \odot \sigma(x) \, , 
	\end{align}
\end{subequations}
where $a$ represents an element-wise nonlinear activation function. Vectors $\mu$ and $\sigma$ parameterize an input-dependent normal distribution in the latent space. The output $z$ is sampled accordingly, using $\epsilon$, a vector that is sampled from a standard normal distribution, and the Hadamard product $\odot$. 
Mirroring the encoder network, the decoder maps the sampled latent space code $z$ to the $d$-dimensional data $\mu'$ and $\sigma'$ using
\begin{align}\label{eq:mu_2}
	\left(\begin{matrix} \mu' \\ \sigma' \end{matrix}\right) = & \left(\begin{matrix} W^{\mu'} \\ W^{\sigma'} \end{matrix}\right)(\ldots a(W_{1}^{d}z + b_{1}^{d})\ldots) + \left(\begin{matrix} b^{\mu'} \\ b^{\sigma'} \end{matrix}\right)\, ,
\end{align}
where $W_{l}^{d}$ and $b_{l}^{d}$ denote weight matrices and bias vectors for decoding, respectively. $\mu'$ and $\sigma'$ parameterize a $z$-dependent normal distribution in the $x$ space. 

After the training process, only the decoder part of the trained VAE network is utilized to generate data. Latent space codes $\Tilde{z}$ are sampled from the standard normal distribution $\mathcal{N}(0,I)$ (see Fig.~\ref{fig:VAE} (b)). Then, data space samples $\Tilde{x}$ are sampled from distribution $\mathcal{N}(\mu'(\Tilde{z}),\sigma'(\Tilde{z}))$ as $\tilde{x} = \mu'(\Tilde{z}) + \epsilon \odot \sigma'(\Tilde{z})$, whose parameters are determined by $\Tilde{z}$. 

\subsection{Training Objective} \label{sec:naive_VAE_function}
During training, weight matrices $W$ and bias vectors $b$ are updated iteratively to minimize the loss function \cite{Kingma2019}
\begin{align}\label{eq:loss_sum}
    \mathcal{L}=\mathcal{L}_{D_{KL}}+\mathcal{L}_{Re}.
\end{align}
The \emph{Kullback-Leibler loss} $\mathcal{L}_{D_{KL}}=\sum_i D_{KL}(q_\phi(z|x_i)||p(z))$ is the sum over all training data points $x_i$ (assumed i.i.d.) of the Kullback–Leibler divergence between that point's posterior distribution $q_\phi(z|x_i)$ and the prior distribution $p(z)$ (chosen as the standard normal distribution). The Kullback-Leibler loss is computed as
\begin{align}\label{eq:KL_loss}
    \mathcal{L}_{D_{KL}}
    & = \frac{1}{2}\sum_{i=1}^n\sum_{j=1}^d(-1+{\sigma^2_{i,j}}+{\mu}^2_{i,j}-\log{\sigma^2_{i,j}}),
\end{align}
where $n$ denotes total number of observations used for training and $(\mu_i, \sigma_i)$ are evaluated for data point $x_i$.
The \emph{reconstruction loss} $\mathcal{L}_{Re}$, representing the negative log-likelihood of reconstructing the inputs $x_i$ via their latent space codes and the decoder that is parameterized by $\theta$, is written as $-\sum_{i=1}^{n} \mathbb{E}_{Z\sim q_\phi(z|x_i)}[\log_{P_\theta}(x_i|Z)]$. With a constant $\frac{n d}{2}\log2\pi$ omitted, $\mathcal{L}_{Re}$ is computed as an expectation approximated by a single-point draw:
\begin{align}\label{eq:Re_loss}
    \mathcal{L}_{Re}  \approx \frac{1}{2}\sum_{i=1}^n\sum_{j=1}^d((x_{i,j}-\mu'_{i,j})^2/\sigma'^2_{i,j}+\log\sigma'^2_{i,j}).
\end{align} 

During training, the full-sample sum in loss functions $\mathcal{L}_{D_{KL}}$ and $\mathcal{L}_{Re}$ are replaced by sample averages. Also, a weighing factor $\beta$ is introduced for $\mathcal{L}_{D_{KL}}$ to adjust the ratio between two losses in \eqref{eq:loss_sum} (known as $\beta$-VAE \cite{burgess2018understanding}):
\begin{align}\label{eq:loss_sum_adj}
    \mathcal{L}=\beta\mathcal{L}_{D_{KL}}+\mathcal{L}_{Re}.
\end{align} 
\begin{figure} 
  \centering 
    \includegraphics[scale=0.8]{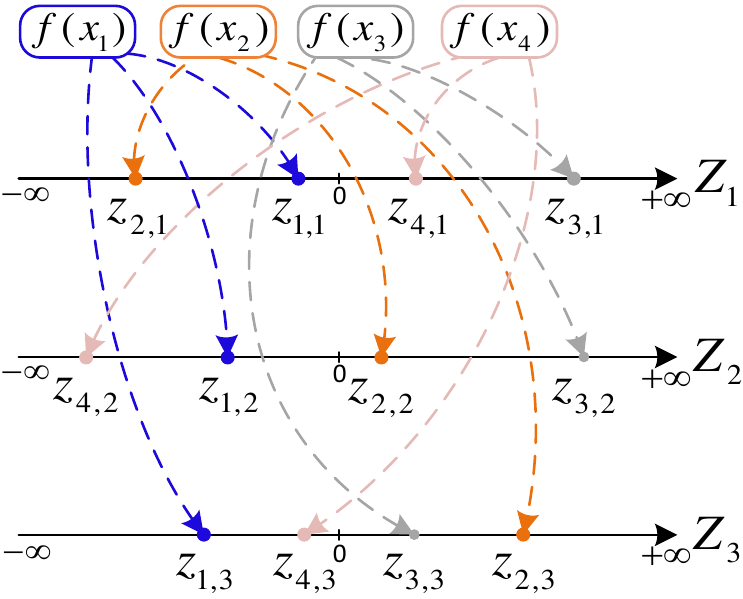}
  \caption{Illustration of disordered latent space code.}
  \label{fig:Schematic_Dis_Order}  
\end{figure}
\begin{figure} 
  \centering 
    \includegraphics[scale=0.8]{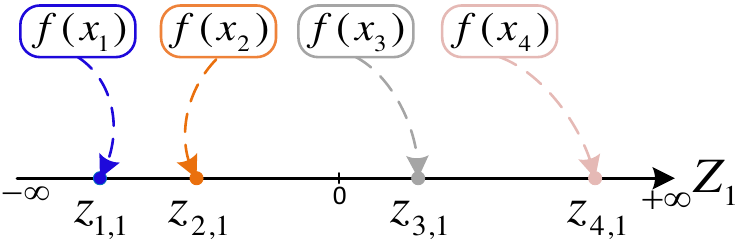}
  \caption{Illustration ordered latent space code.}
  \label{fig:Schematic_Order}  
\end{figure}

\subsection{Proposed Oriented VAE-Based Generative Model} \label{OVAE_model}

For the basic VAE generative model, the distribution of latent space codes approximately follows a multivariate standard normal distribution \cite{Kingma2019}, i.e.
\begin{equation} 
    \frac{1}{n}\sum_{i=1}^n q_\phi(z|x_i) \underset{\mathrm{approx.}}{\sim} \mathcal{N}(0,1)
\end{equation}
However, the relation between codes (values of $z$) and the corresponding states (values of $x$) or features of interest is otherwise unconstrained. If one aims to perform targeted sampling of states, the location of interesting states in the latent space may not be known, or such states may be distributed in ways that prohibit efficient sampling.

In what follows, $f(x)$ indicates a (single, continuous) feature of interest that can be calculated from the state $x$. For risk assessment applications, it may be the state impact ($f(x)=h(x)$) or some approximation thereof. In any case, it is assumed to be sufficient to guide targeted sampling of $x$. Fig.~\ref{fig:Schematic_Dis_Order} illustrates how four different states ($x_1,\dots,x_4$, in order of increasing $f(x)$) may be mapped onto a three-dimensional latent space. In this case, there is no easily apparent sampling strategy that preferentially targets samples with a particular range of $f(x)$.

\begin{figure} 
  \centering 
    \includegraphics[scale=0.65]{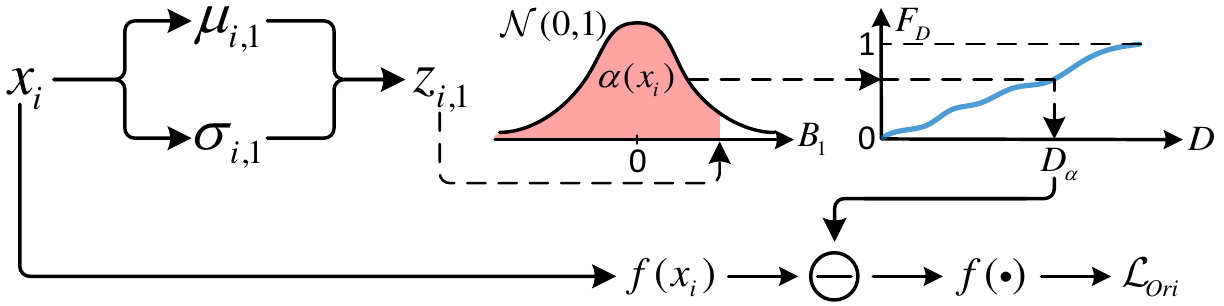}
  \caption{Data flow of calculating $\mathcal{L}_{Ori}$.}
  \label{fig:Data flow}  
\end{figure}
\begin{figure} 
  \centering 
    \includegraphics[scale=0.595]{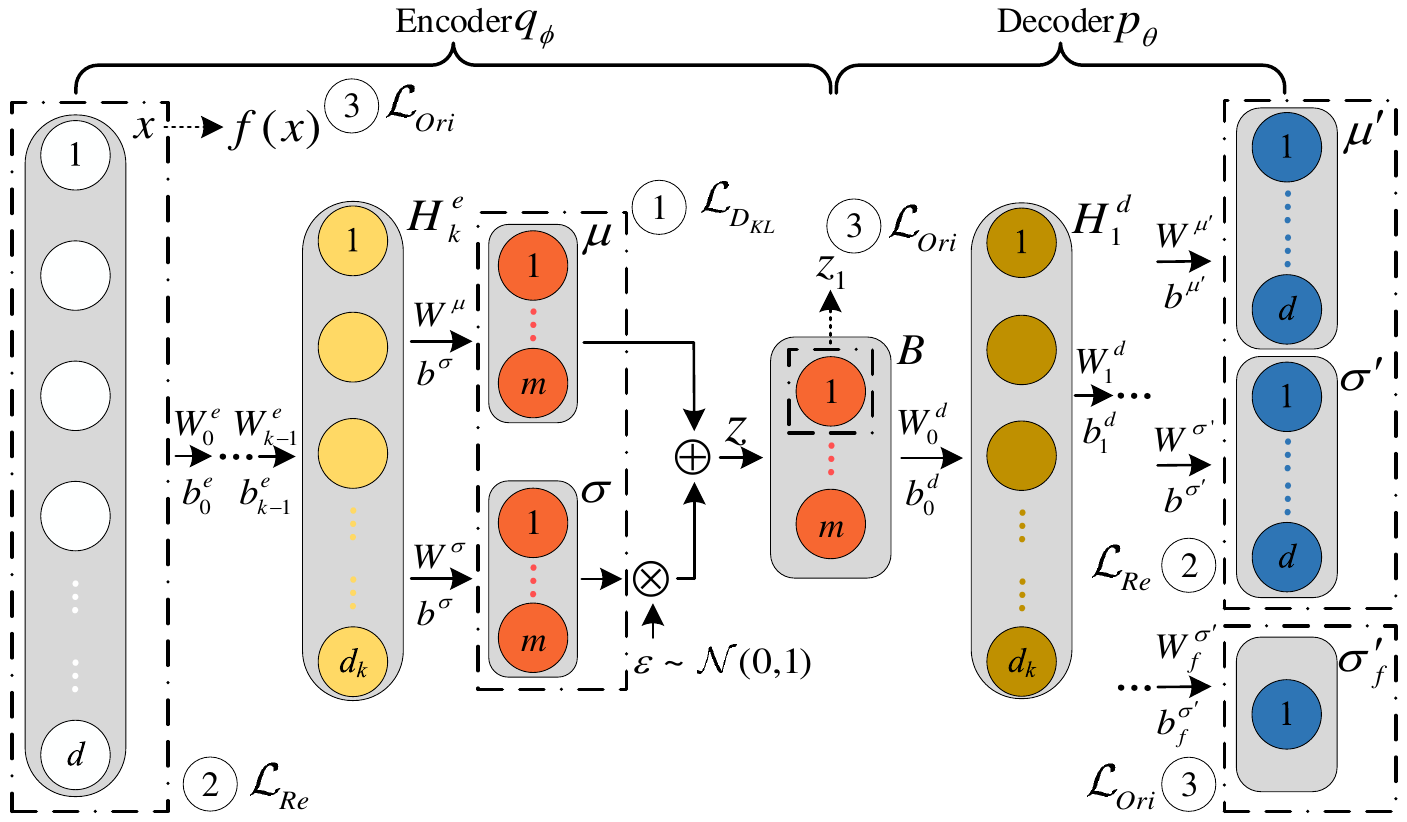}
  \caption{Schematic of the proposed OVAE model.}
  \label{fig:OVAE}  
\end{figure}

Given this, we propose the \emph{oriented variational autoencoder} (OVAE). The idea of OVAE is to force one dimension of the latent space code $z$ to correlate with $f(x)$, while still approximately following a standard normal distribution. 
In this work, we arbitrarily align increasing values of $f(x)$ with $z_1$, the first coordinate of the latent space. In the example given above, the aim is to assign $z_1$ coordinates of samples in order of increasing values $f(x_i)$, as illustrated in Fig.~\ref{fig:Schematic_Order}.

The training process is designed not to disturb the overall distribution of states in the latent space. After training, the known distribution of $z_1$ and its known alignment with $f(x)$ can be used to perform targeted sampling. The remaining coordinates $z_{l\neq 1}$ can be sampled independently from standard normal distributions as for the regular VAE model. 
The change in the training process will be achieved by introducing an additional orientation loss term to the training loss \eqref{eq:loss_sum_adj}:
\begin{align}\label{eq:loss_sum_new}
    \mathcal{L}=\beta\mathcal{L}_{D_{KL}}+\mathcal{L}_{Re}+\mathcal{L}_{Ori}.
\end{align}
The functional form of $\mathcal{L}_{Ori}$ is determined as follows. 
Conceptually, the desired orientation is achieved by considering the value $f(x)$ as an additional feature that must be reconstructed by the autoencoding network during training, according to the following rules: 
\begin{enumerate}
    \item The reconstructed feature (corresponding to a single latent code $z$) $\hat{f} (z(x_i)) \approx f(x_i)$ is normally distributed, i.e. $\hat{f}(z(x_i)) \sim \mathcal{N}(\mu'_f(z(x_i)),\sigma'_f(z(x_i))$.
    \item To ensure the desired association of $z_1(x)$ and $f(x)$, only $z_{1}(x)$ (the first component of the latent code $z(x)$) is used to generate $\mu'_f(z)$. Moreover, this is done using a pre-specified decoder (explained below).
    \item Unlike other features, the value $f(x)$ is not used in the encoder, to facilitate semi-supervised learning (explained below).
\end{enumerate}
The decoding function $\mu'_f(z_1)$ is defined as the idealized monotonic mapping from $z_1$ to $f(x)$. This can be computed \emph{a priori}, by considering that (1) the distribution of $z_1$ over all samples is a standard normal distribution and (2) using the empirical distribution $\hat{F}(x)$ of $f(x)$ as the target distribution. Using the probability integral transform results in 
\begin{equation}
    \mu'_f(z_1) = \hat{F}^{-1}(\Phi(z_{1})),
\end{equation}
where $\Phi$ is the CDF of the standard normal distribution. This procedure is illustrated in Fig.~\ref{fig:Data flow}. Following the steps that led to \eqref{eq:Re_loss} we finally obtain the loss contribution
\begin{align}\label{eq:Ori_loss_adj}
    \mathcal{L}_{Ori}  =\sum_{i=1}^n \left\{ \left[f(x_i)-\hat{F}^{-1}(\Phi(q_\phi(z_{i,1}|x_i)))\right]^2/\sigma'^2_{f_i}+\log\sigma'^2_{f_i} \right\}.
\end{align}
The resulting schematic of the OVAE model is shown in Fig.~\ref{fig:OVAE}. The coordinate $z_{1}$ in the latent space is indicated by the dashed box in the bottleneck layer $B$. Note that, when utilized as a generator, the structure of the OVAE model is the same as that shown in Fig.~\ref{fig:VAE} (b), so the illustration is omitted.

When data labels $f(x)$ are hard-to-get or when the analysis of data labels is time-consuming, data sets may be partially labeled. In this case, the OVAE model can also be trained in a \emph{semi-supervised} manner. When a training batch contains both labeled and unlabeled data, the \emph{Kullback-Leibler loss} $\mathcal{L}_{D_{KL}}$ in \eqref{eq:KL_loss} and \emph{reconstruction loss} $\mathcal{L}_{Re}$ in \eqref{eq:Re_loss} are calculated as an average over all data in the batch, but the orientation loss $\mathcal{L}_{Ori}$ shown in \eqref{eq:Ori_loss_adj} is averaged only over the labeled data.

\subsection{Importance Sampling with OVAE} \label{sec:IS-OVAE}

The OVAE model for data synthesis is highly suitable for use in importance sampling, because it has a known sample distribution in the latent space, and one of the latent space variables encodes a feature of interest. 
In this work, we use a simple importance sampling procedure that makes use of these properties. The biased sampling distribution in the latent space is given by 
\begin{subequations}
\label{eq:all-is-density}
\begin{align}
    \label{eq:g}
    q(z_1) &= \alpha \mathcal{N}(z_1; 0,1) + (1-\alpha) \mathcal{N}(z_1; \mu_{IS},\sigma_{IS}^2), \\
    q(z_{l \neq 1}) &=  \mathcal{N}(z_l; 0,1),
\end{align}
\end{subequations}
where $\mathcal{N}(z; \mu,\sigma^2) = \exp(-(z-\mu)^2/(2\sigma^2))/(\sigma \sqrt{2\pi})$. In this model, the first component $z_1$ is sampled from a mixture of a standard normal distribution and a normal distribution with tunable parameters. The other components are sampled from the standard normal distribution. The importance sampling weights are therefore given by
\begin{equation} \label{eq:is-weight}
    w(z) = \frac{p(z)}{q(z)} = \frac{\mathcal{N}(z_1; 0,1)}{\alpha \mathcal{N}(z_1; 0,1) + (1-\alpha) \mathcal{N}(z_1; \mu_{IS},\sigma_{IS}^2)}.
\end{equation}
Exploding weights for relevant states are a known pitfall of importance sampling methods. As can be seen from \eqref{eq:is-weight}, the hyperparameter $\alpha$ ensures that the sampling weight never exceeds $1/\alpha$. A value of $0.1$ is used in this work, sacrificing a small fraction of the sample in exchange for additional robustness. 

The parameters $\mu_{IS}$ and $\sigma_{IS}$ are chosen such that the importance sampling distribution approximates the optimal sampling distribution \eqref{eq:is-optimal}. The procedure for training the OVAE model and the IS parameters for risk assessment (as described in Section~\ref{IS_for_Risk}) is as follows:
\begin{enumerate}
    \item Define a feature $f(x)$ that equals or approximates the sample impact $h(x)$.
    \item If calculating $f(x)$ is straightforward, label all data points $x_i$. If not, label a random subset.
    \item Train the OVAE model on all data (fully or partially labeled).
    \item Use the expectation-maximization (EM) algorithm to optimize $\mu_{IS}$ and $\sigma_{IS}$ so that \eqref{eq:g} approximates $q^*(z)$.
    \item Sample $\tilde{z}$ according to \eqref{eq:all-is-density}, decode latent samples using \eqref{eq:mu_2} (inserting output noise in the process) and estimate the risk $r$ using \eqref{eq:is_risk} and weights \eqref{eq:is-weight}.
\end{enumerate}

\section{Case Study Description}{\label{sec:European load data}}
The OVAE model was trained to generate snapshots of European country-level electricity demand. The generated samples were then used in a multi-area resource adequacy study, where the ability to generate targeted samples was used to greatly increase the sample efficiency. The study case is similar to the one presented in \cite{wang2022generating}. The data and models used are explained in detail below. 

\subsection{Electricity Demand Data and OVAE Model Structure}

Historical hourly load demand data for 34 European countries from 2017 and 2018 were obtained from the Open Power System Data platform \cite{Muehlenpfordt2019} (package version \emph{2019-06-05}). The columns of CS (Serbia and Montenegro), IS (Iceland), and UA (Ukraine) were dropped due to incomplete records. Moreover, CY (Cyprus) was omitted due to its lack of connection with power systems from other countries. The historical data were randomly split into training and test sets in blocks of one week with the proportion of 4:1 (13,270 training and 3,212 test samples). The training set was min-max normalized before being fed into the OVAE model, and the inverse transformation was applied to generated samples. The total volume of the generated data was the same as the historical training data set, in order to balance them for visual and statistical analysis. However, we emphasize that the purpose of constructing such a generative model is to have the ability to precisely generate limitless non-repeating data according to users' interests, e.g. to reduce the risk of overfitting in downstream machine learning tasks.

The parameters of the generative models were tuned for optimal performance. The network contained 3 hidden layers in the encoder with dimensions all set as 1,000; the bottleneck layer had 4 nodes (4-dimensional latent vector). The decoder also had 3 hidden layers with the same dimensions as the encoder. The ReLU activation function was used, except for the generation of ($\mu$, $\sigma$) and ($\mu'$, $\sigma'$) leading up to the bottleneck and output layers, respectively. The \emph{adaptive moment estimation} (Adam) weight optimizer \cite{kingma2014adam} was utilized with default settings to iteratively optimize the value of weight matrices $W$ and bias vectors $b$. The batch size was 64, and the learning rate related was $10^{-4}$. Training and data generation of the model was conducted in Python using \texttt{tensorflow} on the Google Colab environment using the GPU option. 

\subsection{Resource Adequacy Model}
\label{sec:model}

The multi-area resource adequacy model represents the network as a directed graph with flow limits. The topology, capacities, and available generation in each node were based on \cite{ENTSO2020}, the 2025 scenario of the ENTSO-E 2020 Mid-term Adequacy Forecast (MAF2020). Net transfer capacities between countries were defined as the summation of transfer capacities between their constituent zones. The released data set does not include generators and unit capacities, so we modeled them as follows. The unit sizes for conventional generators were set on a per-country basis as the closest value under 500~MW that was a divisor of the total capacity of the generators in each country. The assumed unit availability was 80\%, and outages were considered independent. The assumed generating capacity of each country was a summation of conventional generating plus a constant 15\% of nameplate wind power capacity. This model is not intended to be an accurate representation of the European network, but to be representative of the studies that can be carried out using the OVAE generative model.

\subsection{Multi-Area Resource Adequacy Impacts}
\label{sec:structure}

For a Monte Carlo-based resource adequacy assessment, each sampled state $s$ consists of snapshots of generating capacity $\overline{g}_{i}$ and demand $d_{i}$ of each area $i$. Load curtailment $c_{i}$ for each node can be calculated based on the flow constraints and dispatching policy using \eqref{all_objectiveFunction_chap5}. This quadratic problem \eqref{all_objectiveFunction_chap5} with variables $\tilde{c}_i$ (curtailment) and $\tilde{f}_{ij}$ (flows) determines $c_i$. It minimizes the total load curtailments and finds curtailments balance between areas relative to the demand in that area~\cite{evans2020assessing}: 
\begin{subequations} \label{all_objectiveFunction_chap5}
\begin{align}
\label{eqn:objectiveFunction_chap5}
c(\overline{g},d) =\underset{\tilde{f}, \tilde{c}}{\text{arg~min}}  \sum_{i\in \mathcal{N}} \frac{1}{2d_{i}}\tilde{c}_{i}^2 +  \tilde{c}_{i} \\
\label{eqn:consa_chap5}
\underline{f}_{ij} \leq {\tilde{f}_{ij}} \leq \overline{f}_{ij}, & &\forall (ij) \in \mathcal{L} \\
\label{eqn:consb_chap5}
0 \leq \tilde{c}_{i} \leq d_{i}, &&\forall i \in \mathcal{N} \\
\label{eqn:consc_chap5}
d_{i}-\overline{g}_{i} \leq \sum_{j < i} \tilde{f}_{ji}-\sum_{j > i} \tilde{f}_{ij} + \tilde{c}_{i} \leq d_{i} ,&& \forall i \in \mathcal{N}
\end{align}
\end{subequations}
Here, $\mathcal{L}$ and $\mathcal{N}$ are the sets of connections (from $i$ to $j$ with $i<j$) and areas respectively. Constraints on power flow $\tilde{f}_{ij}$ from node $i$ to node $j$ are given in \eqref{eqn:consa_chap5}; \eqref{eqn:consb_chap5} limits curtailment and \eqref{eqn:consc_chap5} enforces flow and generating power constraints. This optimization problem is strictly convex and has a unique solution because the objective function (\ref{eqn:objectiveFunction_chap5}) has a positive definite structure, and the constraints are linear. The python package \texttt{quadprog} \cite{pypi} was used to solve this problem.

For a given sampled state $s=(\overline{g}, d)$, the impact $h(\overline{g}, d)$ is calculated using $c$ according to the metric of interest (LOLE or EENS):
\begin{align}
    h_{EENS}(\overline{g}, d) &= 8760 \times h_{EPNS}(\overline{g}, d),\\
    h_{LOLE}(\overline{g}, d) &= 8760\times \mathbbm{1}_{h_{EPNS}(\overline{g}, d) > 0}, 
\intertext{both using the `power not supplied'}
    h_{EPNS}(\overline{g}, d) &= \sum_{i\in \mathcal{N}} c_i(\overline{g}, d). \label{eq:h-epns}
\end{align}
Note that by summing curtailments over areas, these are whole-system adequacy metrics, instead of (more common) per-area metrics.

The power system is a very reliable system, so draws from $h_{EPNS}(\overline{g}, d)$ are very likely to return zero. It is undesirable to use a feature with such limited information for orienting the OVAE latent space, so we measure the `distance from load shedding' for those states where $h_{EPNS}(s)=0$. This quantity $\Delta$ can be defined as the maximum demand that can be added (in proportion to the base demand $d$) before a shortfall event occurs:
\begin{subequations}
\begin{align}
\label{eqn:linearobjectiveFunction}
\Delta(\overline{g}, d) =\underset{\tilde{f}, \tilde{k}}{\text{max}}  ~\tilde{k}\sum_{i\in \mathcal{N}} d_i \\
\label{eqn:consf}
\underline{f}_{ij} \leq {\tilde{f}_{ij}} \leq \overline{f}_{ij}, & &\forall (ij) \in \mathcal{L} \\
\label{eqn:conse}
d_{i}-\overline{g}_{i} \leq \sum_{j < i} \tilde{f}_{ji}-\sum_{j > i} \tilde{f}_{ij} - \tilde{k}d_{i} \leq d_{i} ,&& \forall i \in \mathcal{N}
\end{align}
\end{subequations}
It is easily verified that this has a (non-negative) solution whenever $h_{EPNS}(\overline{g}, d)=0$. 

The auxiliary feature $f(d)$ that is used to train the OVAE model for demand can now be defined in two steps. First, for each demand $d$, we draw $k=100$ generation states $\overline{g}^{(j)}$, with $j=1,\ldots,100$. Second, we define
\begin{equation} \label{eq:eens-label}
    f_{EENS}(d) = \left\{\begin{array}{ll} 
    8760 \times \overline{h}(d), & \textrm{if } \overline{h}(d) > 0,  \\
    - 8760 \times \min_{j}\Delta(\overline{g}^{(j)}, d) & \textrm{otherwise,}
    \end{array} \right.
\end{equation}
where
\begin{equation}
    \overline{h}(d) =\frac{1}{100}\sum_{j=1}^{100} h_{EPNS}(\overline{g}^{(j)},d).
\end{equation}
This is equal to $h_{EENS}(s)$ averaged over 100 generation states when its value is positive, and provides a continuous extension to negative values when no loss of load state is encountered. This makes $f_{EENS}(d)$ suitable as a feature for OVAE alignment.

\section{Results}

This section describes a number of experiments to test the efficacy of the OVAE model in capturing the data distribution, encoding the feature of interest and potential for importance sampling.

\begin{figure} 
  \centering 
    \includegraphics[scale=0.23]{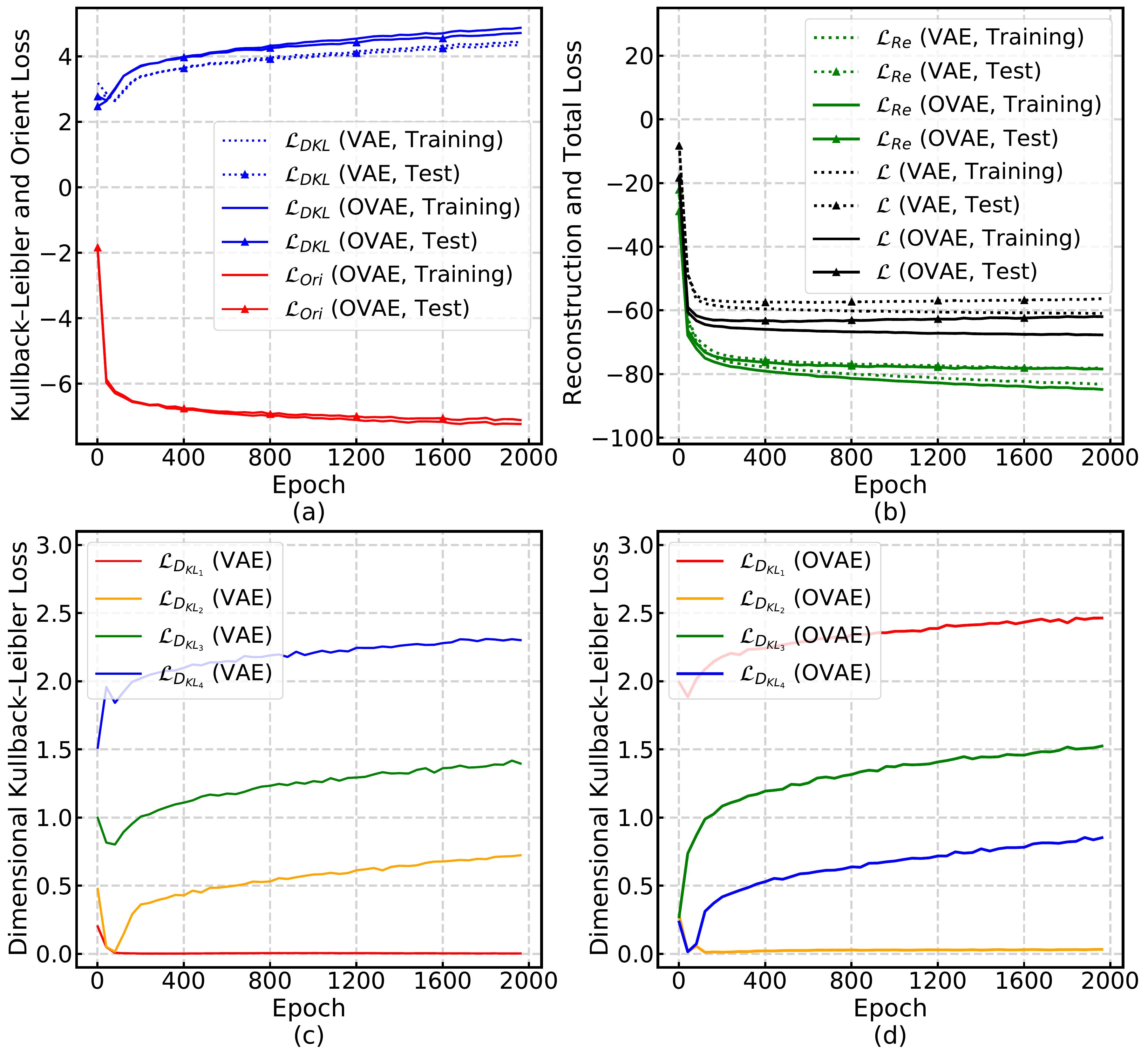}
  \caption{Evolution of loss terms during the training of the VAE and OVAE models.}
  \label{fig:Training process}  
\end{figure}
\subsection{Impact of Extra Oriented Loss $\mathcal{L}_{Ori}$ on Model Training} \label{sec:Impact of L_Ori}

In the first experiment, a simple feature, \emph{total load}, was used to orient the OVAE model. The \emph{total load} 
\begin{equation} \label{eq:total-load}
    f_{TL}(d) = \sum_i d_i
\end{equation}
is readily calculated for each load vector $d_i$ and is expected to be correlated with system stress. Like other training data, this feature is normalized.
The evolution of the loss terms during the training process of the VAE and OVAE models, for $\beta=5$, is depicted in Fig.~\ref{fig:Training process}. The two models show a similar tendency of the \emph{Reconstruction loss} $\mathcal{L}_{Re}$ (Fig.~\ref{fig:Training process}b). The \emph{Kullback-Leibler loss} $\mathcal{L}_{D_{KL}}$ is slightly higher for the OVAE model, probably due to the additional orientation stress imposed by the \emph{Orientation loss} $\mathcal{L}_{Ori}$ (both Fig.~\ref{fig:Training process}a). Detailed comparison of the $\mathcal{L}_{D_{KL}}$ loss for each latent dimension (Fig.~\ref{fig:Training process}c, d) demonstrates that the evolution of the loss (a measure of information contained along its dimension) is similar, but not in identical order. Notably, in the OVAE model, the information content is highest along the first dimension, and for both models, there is an unused dimension ($D_{KL}$=0). For the OVAE model, the total loss $\mathcal{L}$ of the test set starts increasing slightly after 650 epochs. Thus, to avoid overfitting and to compromise on a good training result, we set the number of training epochs as 650.

\begin{figure}[!htp]
  \centering
    \includegraphics[scale=0.23]{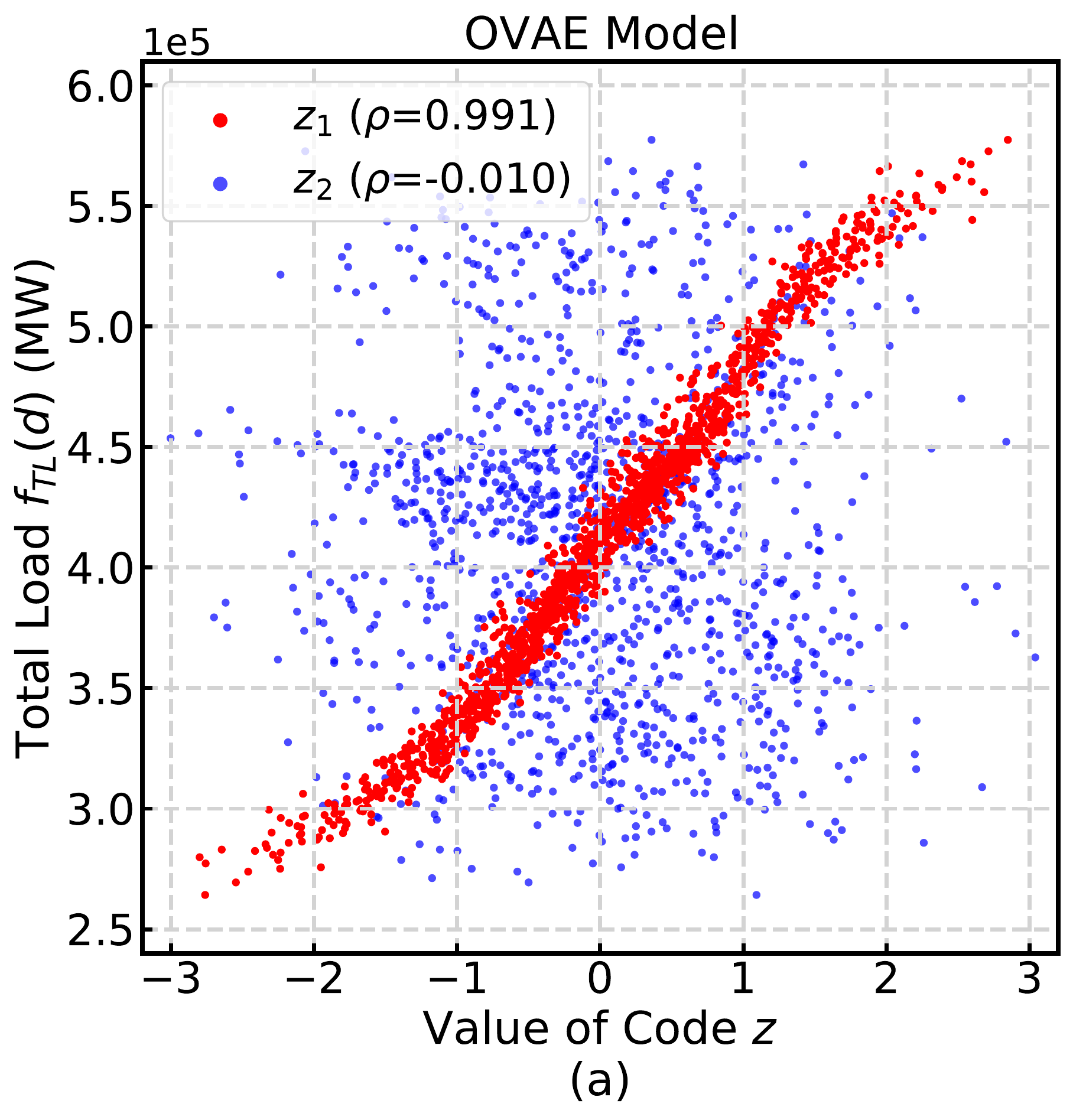} 
    \includegraphics[scale=0.23]{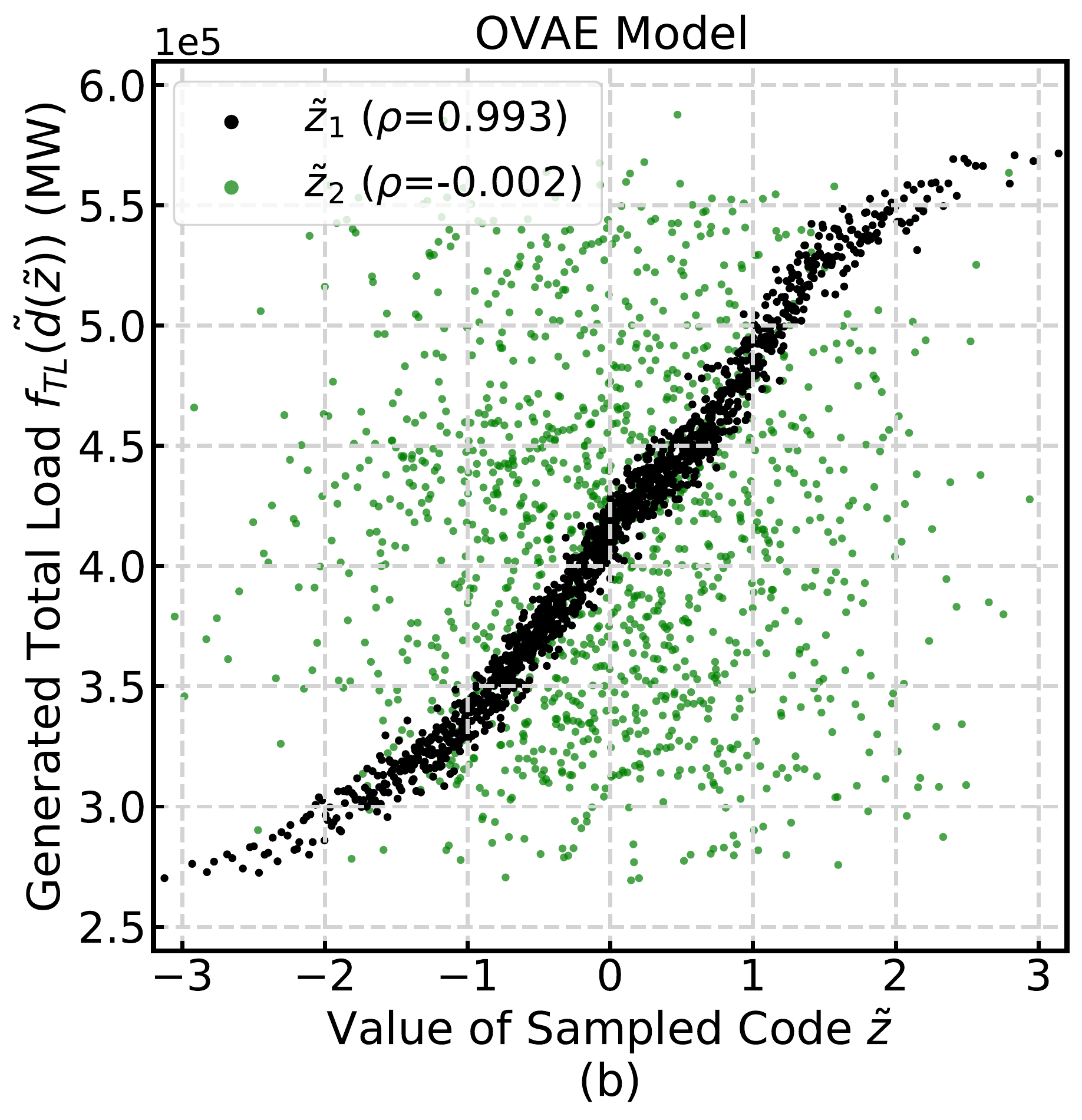} 
    \includegraphics[scale=0.23]{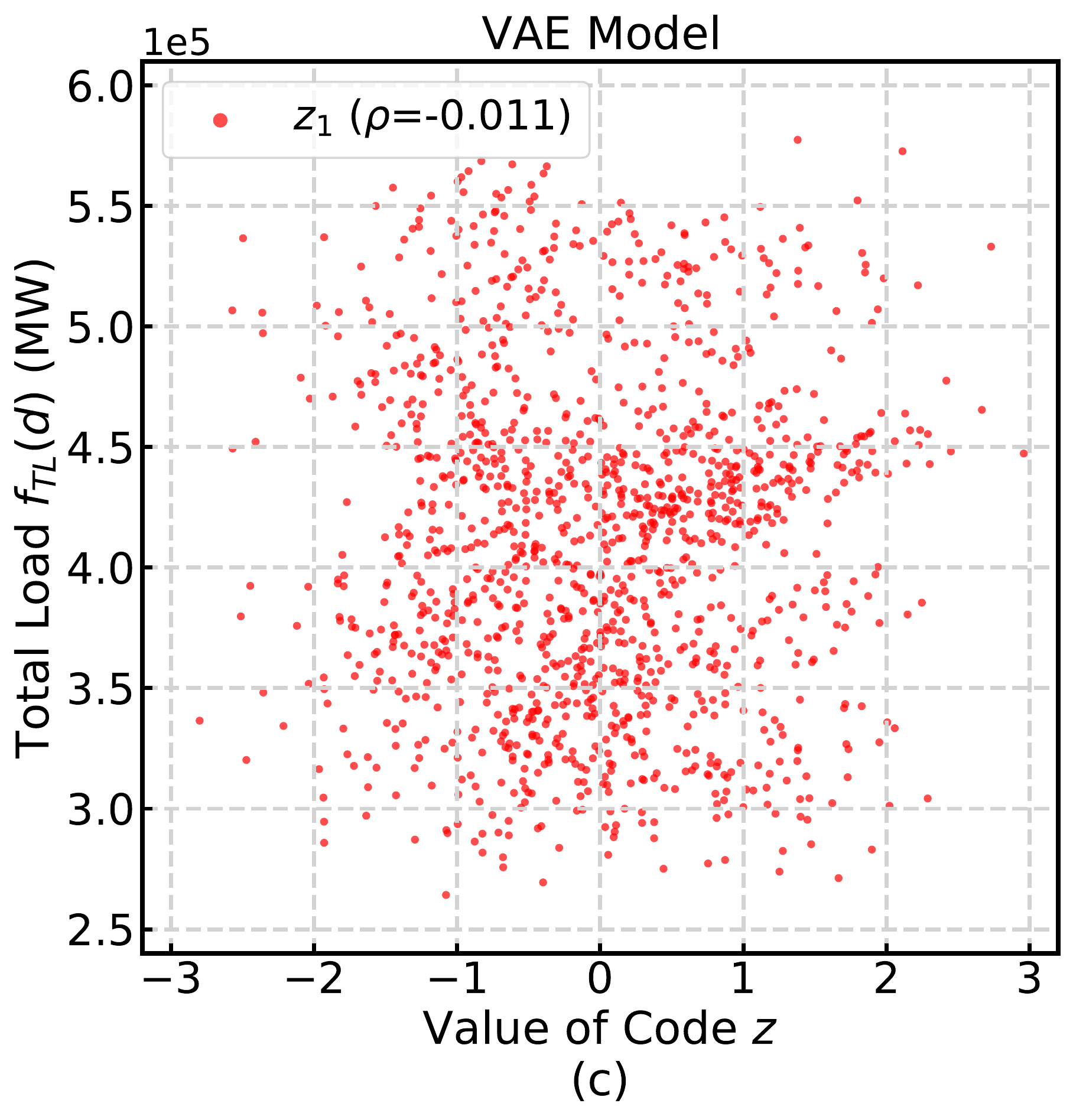}
    \includegraphics[scale=0.23]{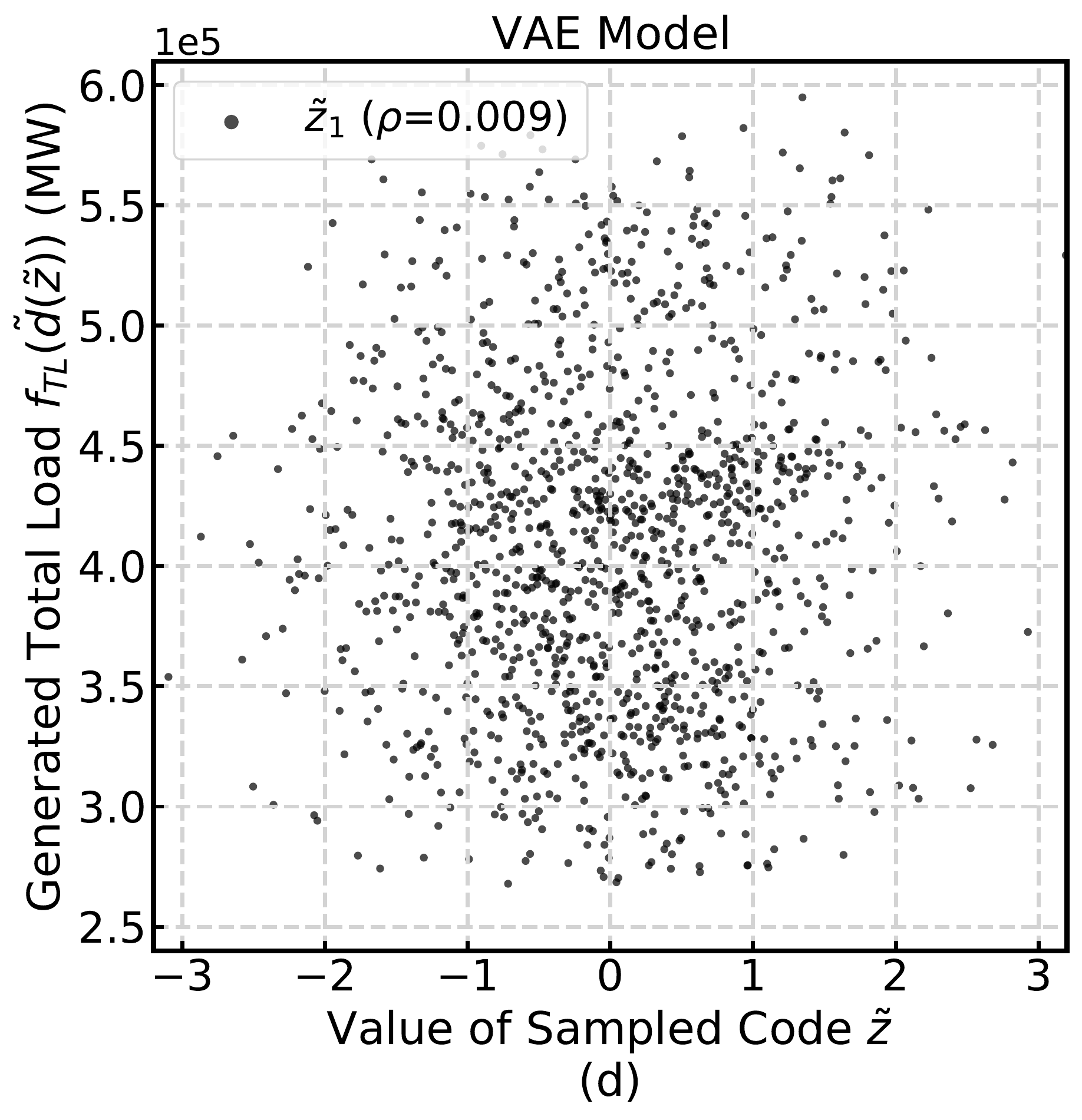}
  \caption{Scatter plot and calculated Spearman correlations between total load (input $f_{TL}(d)$ or sampled $f_{TL}(\Tilde{d}(\Tilde{z}))$) and latent space data ($z$ and $\tilde{z}$) when utilizing OVAE models (a, b) and VAE models (c, d).}
  \label{fig:Correlation}  
\end{figure}
\subsection{Validation of Latent Space Alignment} \label{sec:Correlation}
The second experiment investigated the degree of alignment that is achieved between the feature $f_{TL}(d)$ and the latent code $z(d)$. Fig.~\ref{fig:Correlation}a shows a scatter plot of $z_1(d)$ versus $f_{TL}(d)$ and $z_2(d)$ versus $f_{TL}(d)$, for all training points, using the encoder of the trained OVAE model. The association between $z_1$ and $f_{TL}(d)$ is clearly visible. To quantify this dependence, the Spearman (rank) correlation coefficient was calculated. This association is maintained for \emph{sampled} data: Fig.~\ref{fig:Correlation}b shows the same for samples $\tilde{z}$ that were generated by sampling from the standard normal distribution in the latent space and the total load $f_{TL}(\tilde{d}(\tilde{z}))$ of the reconstructed snapshot. In contrast, no strong correlation between $f_{TL}(d)$ and $z_1(d)$ (or between $f_{TL}(\tilde{d}(\tilde{z}))$ and $\tilde{z}_1$) is present for the VAE model (Fig.~\ref{fig:Correlation}c, d). 

\begin{figure} 
  \centering 
    \includegraphics[scale=0.19]{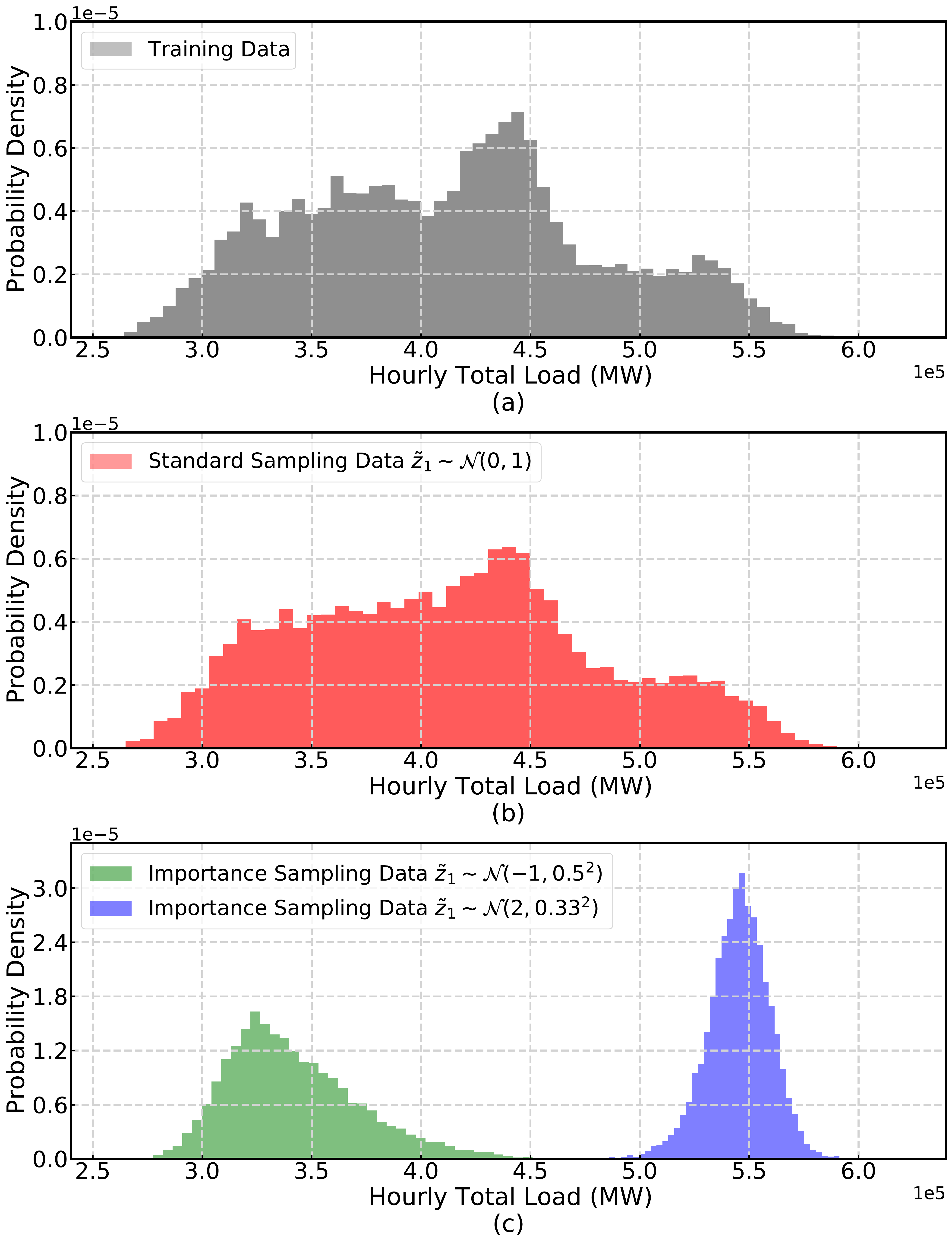}
  \caption{Total load of generated samples, compared between training data (top), unbiased sampling from the OVAE model (middle) and biased sampling (two variations) from the OVAE model (bottom).}
  \label{fig:ResultsOffset Gen}  
\end{figure}
\subsection{Unbiased and Biased Sampling} \label{sec:Nonstandard Sampling}

Apart from sampling $\tilde{z}$ from the standard normal distribution $\mathcal{N}(0,1)$, we changed the distribution of $\tilde{z}_1$ and observed the corresponding distribution of the total load $f_{TL}(\Tilde{d}(\Tilde{z}))$. Note that the other three dimensions of $\tilde{z}$ (i.e., $\tilde{z}_2, \tilde{z}_3$ and $\tilde{z}_4$) were still sampled from the standard normal distribution.
Histograms of total load generated by using standard or nonstandard normal distributed $\tilde{z}_1$ are depicted in Fig.~\ref{fig:ResultsOffset Gen}. When $\tilde{z}_1$ is sampled from $\mathcal{N}(0,1)$, the distribution closely resembles that seen in the historical (training) data. By changing the sampling distribution, targeted generation of low or high load states is possible. 

\begin{figure*} 
  \centering 
    \includegraphics[scale=0.223]{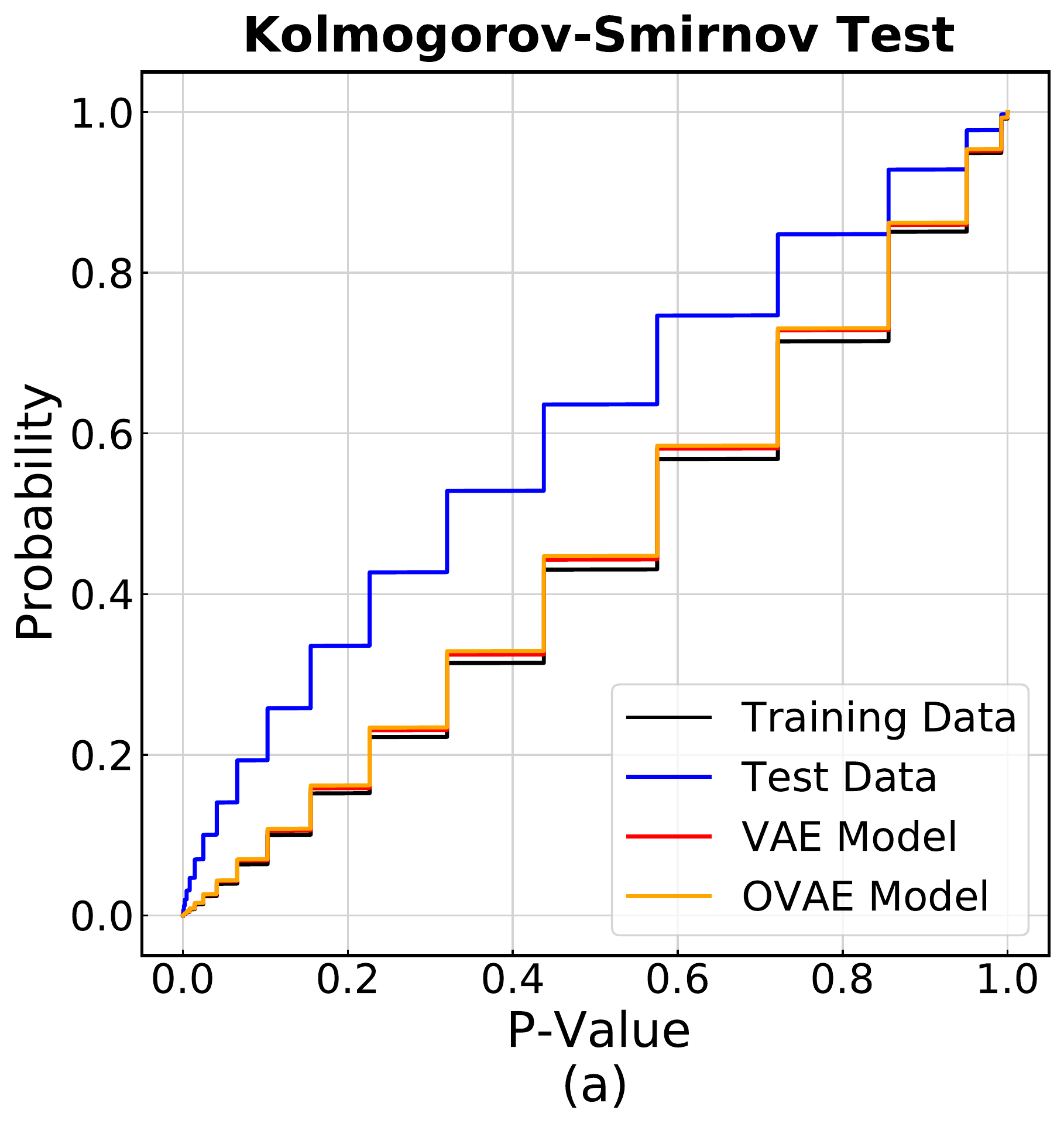}
    \includegraphics[scale=0.223]{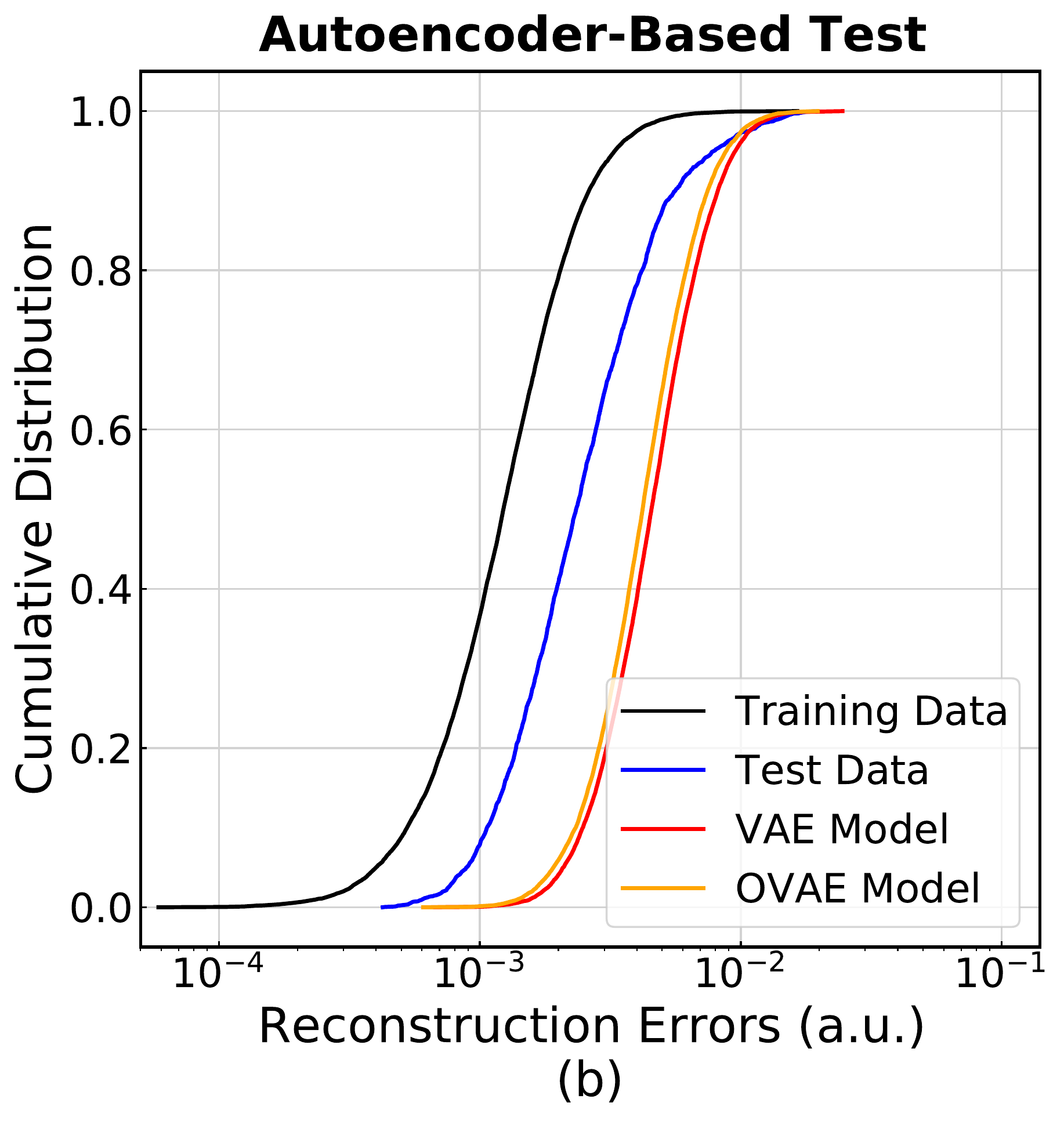}
    \includegraphics[scale=0.223]{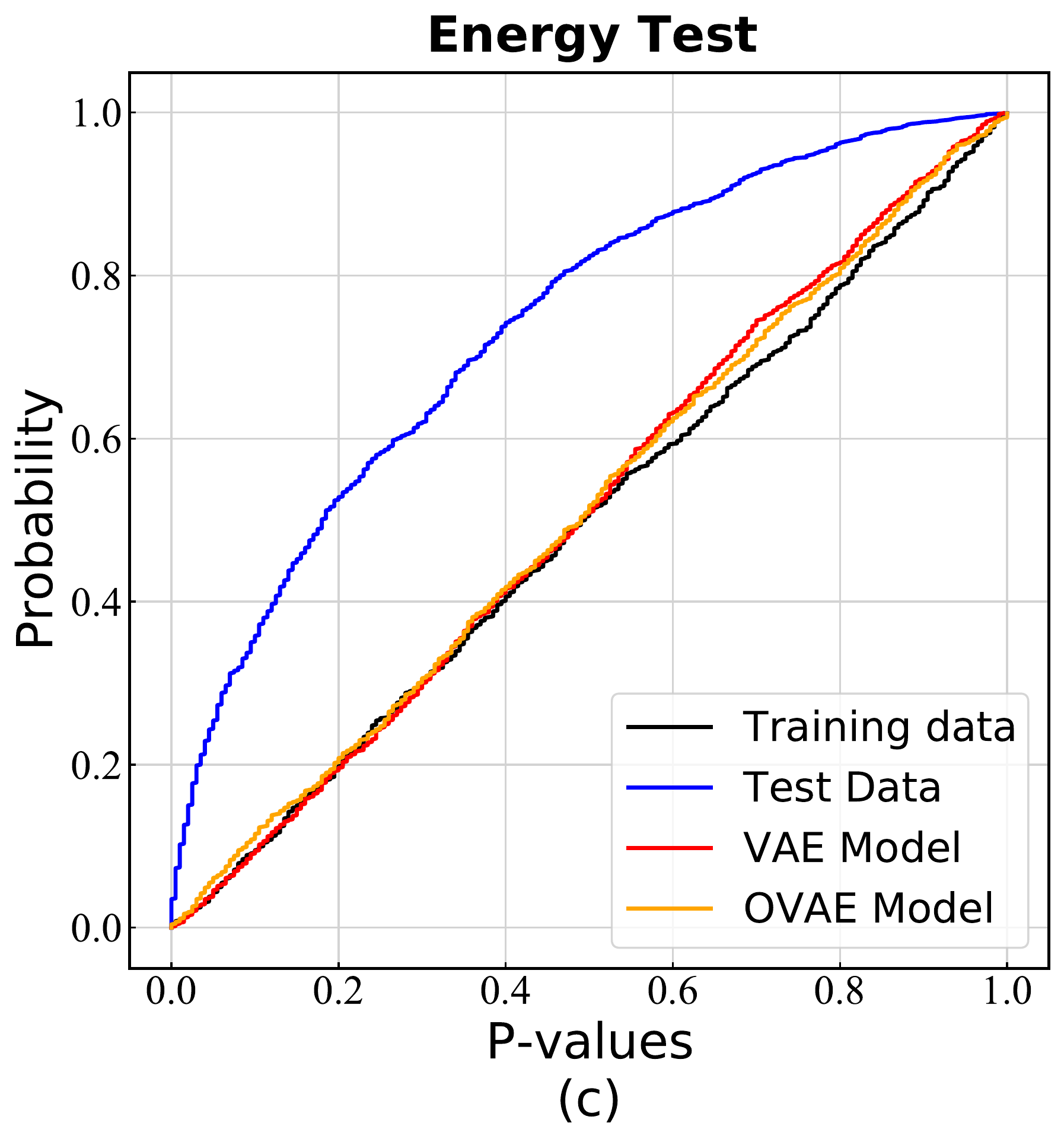}
  \caption{Statistical test results of historical data and generated data. (a) Kolmogorov-Smirnov test. (b) Autoencoder-test. (c) Energy test}
  \label{fig:Statistical test} 
\end{figure*}

\subsection{Quality Evaluation of Generated Data} \label{sec: Data quality}

To further test the capacity of the CVAE model to generate realistic load profiles, non-visual statistical tests were implemented to inspect different aspects of the generated samples. Specifically, in this experiment, the Kolmogorov-Smirnov test, autoencoder-based test, and energy test were utilized to examine univariate marginal distributions, point-wise multivariate dependencies, and multivariate dependencies of population, respectively (see \cite{wang2022generating} for a more extensive explanation). The statistical properties of four models were studied: the OVAE model, the VAE model, random sampling from the training set, and random sampling from the test set.

The Kolmogorov-Smirnov test (Fig.~\ref{fig:Statistical test}a) assesses the accuracy of the marginal distributions. A $p$-value was calculated by comparing 66 random country-level demand values from the training set (0.5\% of the training data set) with the same number of samples from the study model. This was done for each country, and repeated 5,000 times. The results were combined into $p$-value curves for each study model. The experimental results demonstrate a significant difference between the training and test sets. Compared to the test sets, data generated by the OVAE and VAE models have more similar marginal distributions to that of the training set. 

The autoencoder test trains a separate (regular) autoencoder on the training data and tests point-wise multivariate dependencies. The distributions of reconstruction errors obtained using real and generated data are shown in  Fig.~\ref{fig:Statistical test}b. The results indicate that typical reconstruction errors of demand snapshots generated by VAE and OVAE models are larger than those of the reconstruction errors of training and test distributions. A difference between training and test sets is also visible here. 

Finally, the energy test quantifies the similarities of multivariate dependencies of population, compared to the training set. The same as for the K-S test, we used random subsets of 66 data points of the historical and generated population. We used 200 permutations and repeated 1,000 times to draw a distribution of $p$-values. The results in Fig.~\ref{fig:Statistical test}c show that the distribution of samples generated by the OVAE and VAE models is (in this sense) much closer to the training set than that of the test set. 

\subsection{Effectiveness of Performing Semi-Supervised Learning} \label{sec:Semi-Supervised Learning}

In this section, experiments will be conducted to test if our proposed generator can work properly on incomplete labeled data. 
Specifically, the OVAE model was trained on a data set that is partially labeled using the computationally intensive label $f_{EENS}(d)$ \eqref{eq:eens-label} for resource adequacy studies. Different percentages of labeled data were used (5\%, 20\%, 30\%) and labels were replaced by their normalized ranks prior to training. 

\begin{figure} 
  \centering 
    \includegraphics[scale=0.223]{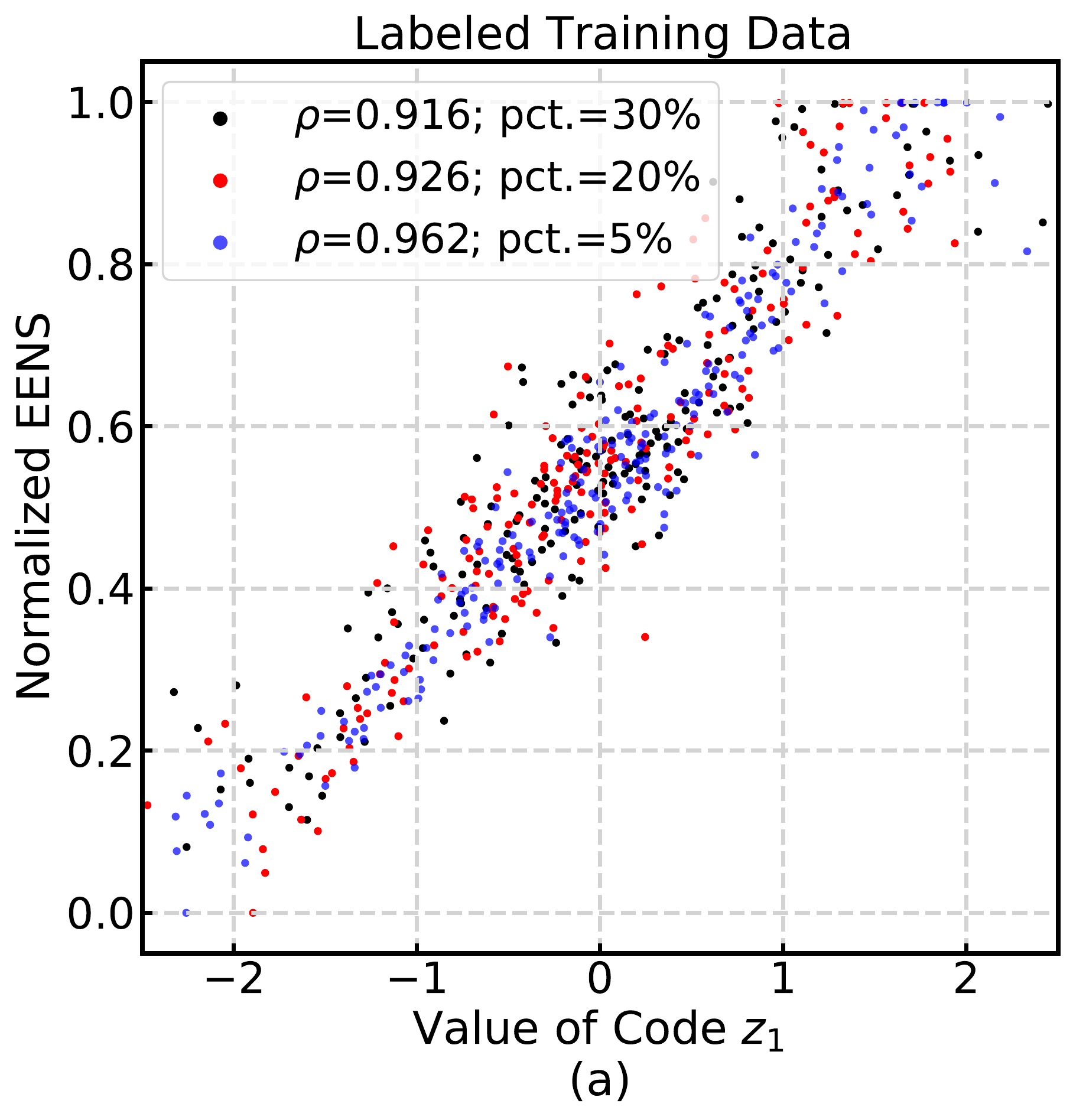}
    \includegraphics[scale=0.223]{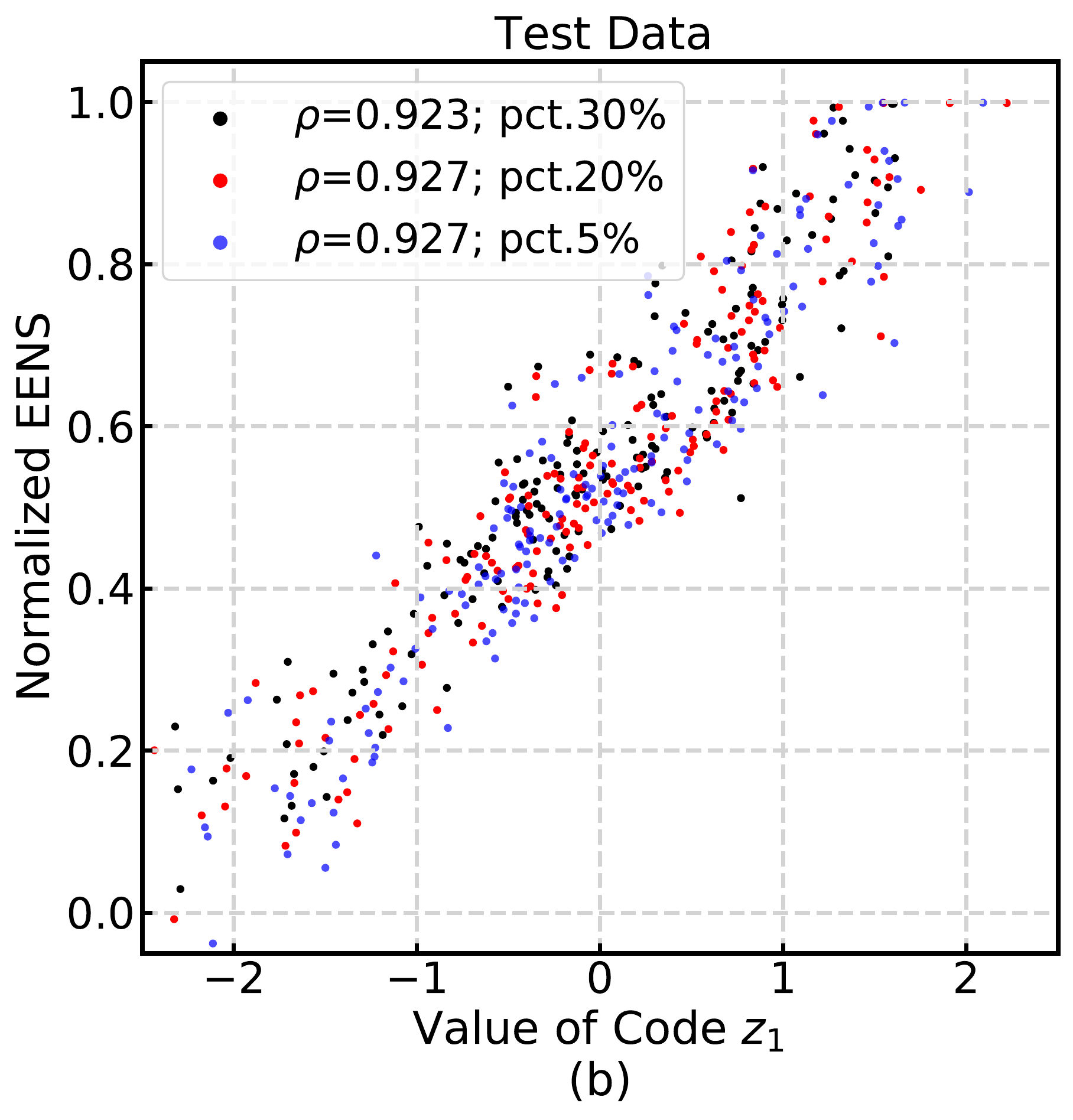}
  \caption{Scatter plot and calculated Spearman correlations between the normalized rank of $f_{EENS}(d)$ and latent space data $z_1(d)$ when trained with different percentages of labeled data.}
  \label{fig:EENS test} 
\end{figure}

The experimentally observed dependencies are shown in Fig.~\ref{fig:EENS test}, for training data (left) and test data (right). 
Using data with different percentages (pct.) of implicit labels for training, $f_{EENS}(d)$ and latent space data $z_1$ show high Spearman correlations. Notably, data with 5\% labels have the highest Spearman correlation, which could be because the small volume of labeled data makes it easier to shape the latent space during training. On the other hand, Spearman correlations are relatively stable on the test data set.

\begin{table*}[h!tb]
    
    \caption{Resource Adequacy Results and Importance Sampling Speedup}
    \begin{tabular}{p{3.5 cm} p{1cm} p{1cm} p{1cm} p{2cm} p{2.3cm} p{1.77cm} p{1.75cm}}
    
    \hlineB{4}
    Load model &$\mu_{IS}$&$\sigma_{IS}$&Time (s)&LOLE (h/y)& EENS (MWh/y)&LOLE Speedup& EENS Speedup\\ \hline
        Historical load& - & - & $4319$ & $10.79(31)$ & $1.190(47)\times10^4$& n/a& n/a\\
        \hline
        OVAE-Total Load &$0$&$1$ &$4155$&$18.76(40)$&$4.50(17)\times10^4$&n/a&n/a\\
        OVAE-EENS 5\% training&$0$&$1$ &$4236$&$18.45(40)$&$3.20(10)\times10^4$ &n/a&n/a\\
        OVAE-EENS 20\% training&$0$&$1$ &$4130$&$18.66(40)$&$3.46(12)\times10^4$  &n/a&n/a\\
        OVAE-EENS 30\% training&$0$&$1$ &$4131$&$18.16(40)$&$2.99(9)\times10^4$ &n/a&n/a\\
        \hline
        OVAE-Total Load &$2.25$&	$0.68$&$4666$&$18.43(20)$&$4.137(40)\times10^4$&$3.5$&$14.5$\\
        OVAE-EENS 5\% training&$2.04$&$0.58$&$4524$&	$18.06(20)$&$3.333(35)\times10^4$&$3.8$&$8.7$\\
        OVAE-EENS 20\% training&$2.00$&$0.48$&$4512$&$18.55(19)$&$3.530(42)\times10^4$&$4.2$&$7.3$\\
        OVAE-EENS 30\% training&$1.92$&$0.60$&$4511$&$18.17(25)$&	$3.217(43)\times10^4$&$2.3$&$5.0$\\    
    \hlineB{4}
    \end{tabular}\label{tab:adequacy}
\end{table*}

\subsection{Multi-Area Adequacy Assessment Results}
Finally, we tested a variety of load models in combination with the resource adequacy model defined in Sections~\ref{sec:model} and \ref{sec:structure}. For load models, we considered historical data and 4 OVAE models that were trained with different settings. One used the total load $f_{TL}(d)$ \eqref{eq:total-load} as a heuristic feature, and the others were based on the more elaborate $f_{EENS}(d)$ feature (\eqref{eq:eens-label}). Different percentages of the training data (5\%, 20\% and 30\%) were labeled.

Simulations to estimate the risk metrics LOLE and EENS were done using $1,000,000$ independent samples of demand and available generating capacity. Simulations were implemented in Python 3.8.12 and were run under Windows 10 x64 on a PC equipped with a 4-core Intel Xeon W-2223 CPU (3600 MHz). 

To estimate the parameters of the importance sampling distribution \eqref{eq:all-is-density}, the procedure described in Section~\ref{sec:IS-OVAE} was followed. The OVAE with standard normal distributions was used to generate 100,000 load states. From this collection of load states, 100,000 states $d^{(j)}$ were drawn at random along with random generating capacity states $\overline{g}^{(j)}$ and the sample weighting function 
\begin{equation}
g(\overline{g}^{(j)},d^{(j)})=\mathbbm{1}_{h_{EPNS}(\overline{g}^{(j)}, d^{(j)}) > 0}
\end{equation}
was used to assign weights to the sampled points. This selects demand states that cause a shortfall, according to the likelihood for this to happen. Finally, expectation maximization was used to estimate $\mu_{IS}$ and $\sigma_{IS}$ in \eqref{eq:g}.

Table~\ref{tab:adequacy} shows the estimated LOLE and EENS values of the European continent for historical load and OVAE models. Risk values are reported in the scientific format, followed by the estimated standard error of the least significant digits in parentheses. For example, $1.190(47)\times10^4$ stands for an estimate of 11,900 with a standard error of 470. When importance sampling was used, the optimized values for $\mu_{IS}$ and $\sigma_{IS}$ are indicated. 

The speedup of sampling-based estimator $A$ with respect to $B$ can be estimated (using the asymptotic speed measure from \cite{tindemans2020accelerating}) as 
\begin{align}
    \label{eq:speedup}
    \textrm{speedup} = \frac{\hat{r}_A^2 t_B \textrm{SE}(\hat{r}_B)^2}{\hat{r}_B^2 t_A \textrm{SE}(\hat{r}_A)^2},
\end{align}
where $t$ is the execution time of simulation, $\hat{r}$ is the estimated value of the risk metric and $\textrm{SE}(\hat{r})$ is its standard error. Estimated speedup values for the LOLE and EENS risk metrics are indicated in Table~\ref{tab:adequacy}.

A few conclusions can be drawn from these results. First, all OVAE models, with or without importance sampling, generate LOLE results that are compatible within their margin of error. However, the LOLE results of around 18 hours per year are all higher than that obtained using the historical load values (11 hours per year). That is not unexpected, given the fact that smooth generative models necessarily extrapolate the historical load distribution and will thus generate more extreme demand values. 

The gap between historical and generative models increases for the EENS metric that is more sensitive to extreme load values. Here, although the OVAE models trained with (partial) $f_{EENS}(d)$ labels offer results that are consistent with each other, the OVAE model trained on $f_{TL}$ labels consistently returns  higher EENS values. This suggests that the use of the $f_{TL}$ label results in a sample distribution that is slightly heavier in the tails, at least in this instance.

Finally, significant speedups are consistently observed when importance sampling is employed, for both the LOLE and EENS metrics. Although not as large as speedups observed for purpose-made importance sampling schemes \cite{da2010generating}, it is important to emphasize that the OVAE is a generic data-driven generative model, that can be deployed in a large variety of situations and levels of modeling complexity.

\section{Conclusion}

In this paper, an oriented variational autoencoder was proposed to relate the first dimension of latent space codes and the characteristics of original space data by Spearman correlation, using an extra orientation loss $\mathcal{L}_{Ori}$ shown in \eqref{eq:Ori_loss_adj}. 
The performance of the OVAE-based data generator was tested by comprehensive experiments. With an extra orientation loss added, the convergence of the other two losses was affected only in a limited way. Moreover, the OVAE generations demonstrated comparable qualities of data points as those generated by the VAE model. When the data set was completely labeled with easy-to-get labels (i.e., total load) that are approximately related to the exact generation goal (i.e., system risk), the experimental results demonstrated not only a good training correlation between data properties and the first dimension codes but also a satisfactory generation correlation between them. Using the trained OVAE model, the generation process of data with user‑defined characteristics was significantly accelerated with well-specified importance weight. Additionally, when only 5\% of the training data was labeled with precise information on system risk (i.e., EENS), the OVAE model was still properly working, showing a high Spearman correlation between data properties and latent space codes as well as speeding up the targeted data sampling process. In the future, we consider embedding physical constraints in data generators to limit the synthetic loads within reasonable physical boundaries.

\bibliographystyle{IEEEtran}
\bibliography{literature}

\end{document}